Research Article

# A FORMAL APPROACH FOR AGENT BASED LARGE CONCURRENT INTELLIGENT SYSTEMS

[1]Chaudhary Ankit, [2]Raheja J L

**Address for correspondence**
[1] Dept. of Computer Science, BITS Pilani, Rajasthan, INDIA-333031
Email: ankit@bits-pilani.ac.in
[2]Digital Signal Processing Group, CEERI, Pilani, Rajasthan, INDIA-333031
Email: jagdish@ceeri.ernet.in

**ABSTRACT**

Large Intelligent Systems are so complex these days that an urgent need for designing such systems in best available way is evolving. Modeling is the useful technique to show a complex real world system into the form of abstraction, so that analysis and implementation of the intelligent system become easy and is useful in gathering the prior knowledge of system that is not possible to experiment with the real world complex systems. This paper discusses a formal approach of agent-based large systems modeling for intelligent systems, which describes design level precautions, challenges and techniques using autonomous agents, as its fundamental modeling abstraction. We are discussing Ad-Hoc Network System as a case study in which we are using mobile agents where nodes are free to relocate, as they form an Intelligent Systems. The designing is very critical in this scenario and it can reduce the whole cost, time duration and risk involved in the project.

## INTRODUCTION

Development teams during the system development life cycle, mostly uses functional analysis and data flow, or object-oriented modeling, which are not sufficient in many cases in themselves, to capture today's dynamic and flexible requirements of some of the current complex projects that are undertaken. Researchers are now seeking new methods and approaches that can help System designers to grapple with some of these problems. One of the new approaches that have been proposed is agent-based modeling. The fundamental notion on which Agent based engineering is autonomous agent. One Key reason to consider an agent as an autonomous system is capable of interacting with other agents in order to satisfy its design objectives, and a naturally appealing one for system designers [1]. Agent modeling in system engineering is a relatively young area, and there are, as yet, no standard methodologies, development tools, or system architectures. System can be defined using multi agents as they can work concurrently to increase the performance of system. A design level strategy is needed to secure the fortune of designed system and to care all the coming problems in advance. Agents also have their several different kinds of problems that need a different kind of treatment. Our work focus is on the design phase for large intelligent systems.

**AGENT-BASED MODELING**

The idea of agents involves from artificial intelligence and neural networks. To show intelligence and self learning mode in agents for self learning Intelligent Systems, these approaches are highly needed. Those behind this movement assert that key techniques for managing complexity, such as decomposition, abstraction, and organization can be comfortably accommodated within the agent-based modeling. A little work has been done in this area by Luck [2], who uses Z formal language for different level agent designing frameworks. Fisher [3] used different logics to check and model for agents and their executions like METATEM. An agent can be treated as a human agent that works on behalf of a person and can represent the person at some places. In systems also if human interaction is not available then also intelligent agent can perform the functions according to the environment changes. It may be predefined or can be perform by agent by self learning. It can be called about the agent that intelligent behavior is the selection of actions based on knowledge. Agent can be destructive also like computer virus [4], if modeled badly. Consequently the promoters of Agents argue that agent-based approaches can significantly enhance our ability to model, design and build complex systems, as they provide different flexibilities. Agent-based techniques can be implemented during the design phase, where agent-based techniques can be used to model the problem domain and the system design, and during the implementation phase, where agent-based development tools could be used. The use of the agent based approach in both phases is a natural fit, but not required. The use of agent-based methodology is increasing rapidly in industrial sector, for simulated applications, motor control, Sensor specific applications and other critical systems, this is beneficial. However, this is not necessary to use both design approaches (Object based Vs Agent Based) together; it would fail to take advantage of the natural mapping from one phase to the other. People like Hilaire [5] have tried to link formal approaches and agent modeling. Here we will make reference to agent-based design and our work will concentrate on agent-based development frameworks for Intelligent Systems.

## AGENT VS OBJECT MODELED DESIGN

An intelligent system which makes the life simpler like intelligent washing machine or microware oven, are harder to build than other simple same kind of system. In certain domain of problems they are very critical and time bounded, so that they become very complicated. Domain like telecommunications, robotics control, manufacturing systems, intelligent systems have different perspectives , so their designing is very critical and all system success is dependent on  the designing of system. As the easiness of the system increases, the complexity of the implementation increases. The design itself is very complicated. People think about object-based design, there are certain difference between object- based and agent-based design.

1. Objects are passive in nature, without invocation message, mostly they never  active [4], that is against the nature of Intelligent systems.
2. In object orient methods action choice is not defined, any member can invoke any publicly available object [4], while in Intelligent Systems, system have to take action according to the environmental changes.

Agent based systems are very complicated in nature. On the other hand object-based does not provide a good set of methodology to form a model of these systems. So to make easy understandability and strong high level design the component architecture is used. Agents are the computer software that functions autonomously without human intervention. Agent-based programming is the extension of object-based programming with some improvement. The idea behind agent-based systems, are that they are capable to reconfigure or operate themselves whenever they needed. OO methodologies are not directly applicable to agent systems. An Agent based Models of G-Nets are shown in Fig 1. Agents are usually significantly more complex than typical objects, both in their internal structure and in the behaviors they exhibit [6].

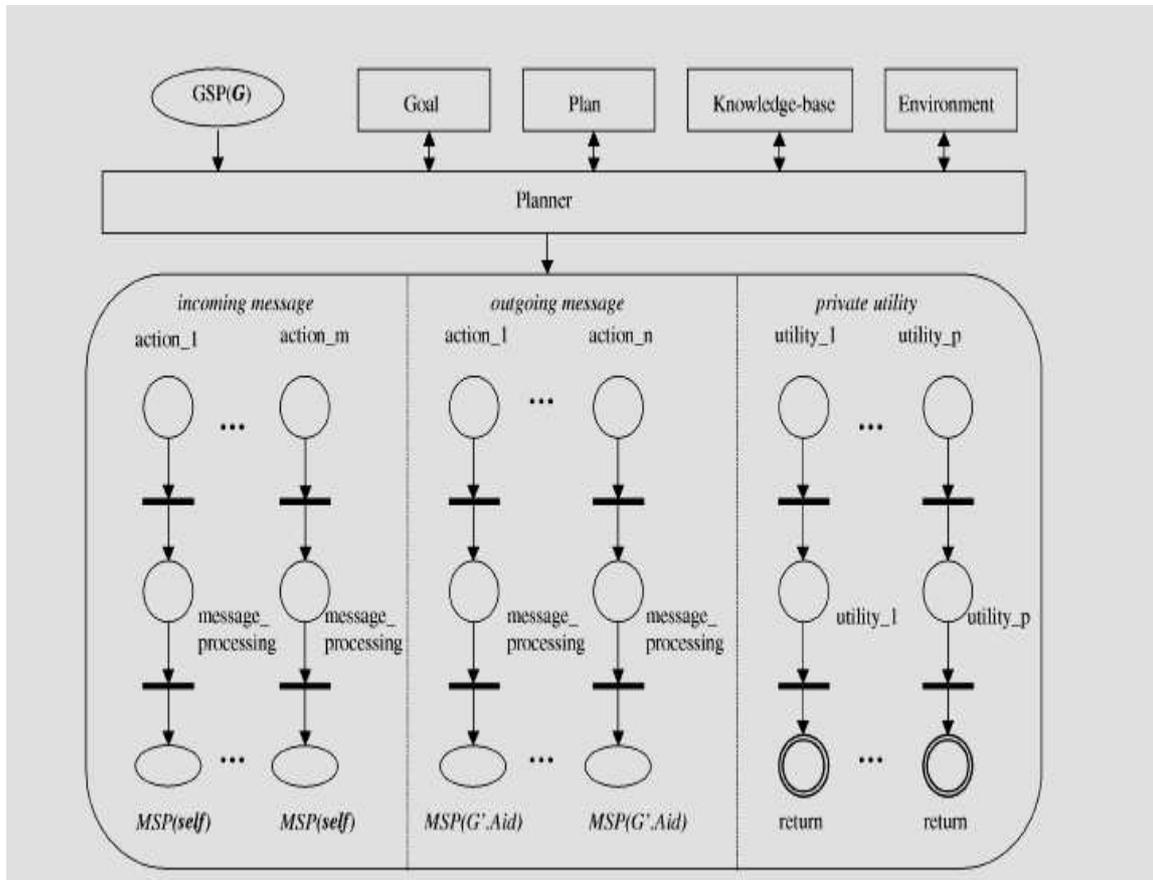

**Figure 1. A Generic Agent Based Modeling for G-Net [1]**

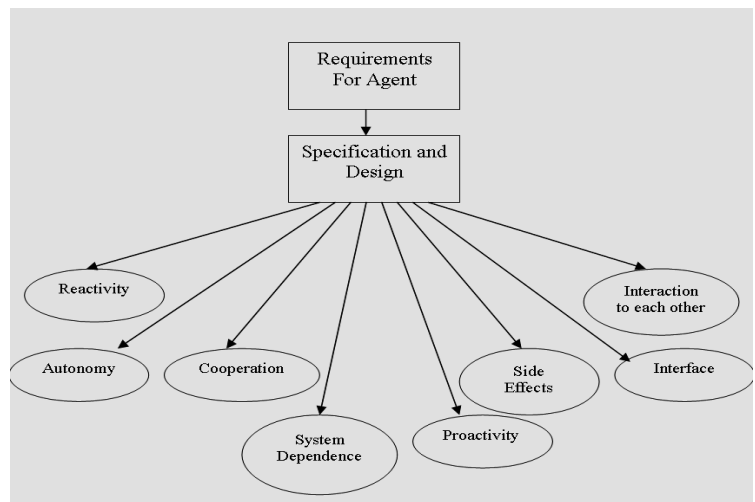

**Figure 2:. Issue in Agent Based Modeling Software System**

International Journal of Advanced Engineering Technology

## ISSUES IN AGENT DESIGN

The agent paradigm is based upon the notion of reactive, autonomous, internally-motivated entities embedded in changing, uncertain environments which they perceive and in which they act [6]. Traditional system engineering approaches offer limited support for the development of intelligent systems. To handle the tremendous complexity and the new engineering challenges presented by intelligence, self learning, adaptive ness and seamless integration, developers need higher-level development constructs [7]. The Complexity of the system is obvious because synchronization and interaction (called as *meet* in agent modeling [8]) among the agents and different functionality of the system are tedious jobs to do. Although we know many techniques from software engineering but still a better one is needed. There can be several issues in the agent based system design, but most of them will be system dependent that is applicable to only that particular application domain. Here we are discussing main issues or say properties of agents that should maintain during the design and should continue till the implementation. These issues are above the design techniques [9]. We recommended designers to use modular and abstraction approach to reduce complexity. Hierarchical decomposition is also possible that depend on system. Agents are entities that have predefined goals, mechanically they triggered either from internal or external change and work according to it. Few properties of Agents are-

- During the operation, Agents perform so many action and reactions but when they get accessed from outside the world they come up with a known state. That is state of that agent.
- Agents are defined in the system to fulfill the system goal. They have their internal sub goals that they attempt to fulfill though their actions in the process. These sub goals are predefined in the system to access it or to do function according to them.
- Agents have the goals that are part of system goal. Agent can start some action to fulfill the goals, called pro-activity. This is the action in advance, based on some information.
- Agents are an intelligent entity, so it can operate itself without the user interrupt. This property is autonomy.

- Agents operate without any direct input and have control over their actions and internal state.
- Agents have to interact to each other for different purposes, several times for it fulfilling the goal and several times to fulfill the goal of the system. *Meets* have a great role in agent based system modeling. Proper communication behavior of agents and interface, connectors in the system through which they will communicate should be well defined.
- Agents are the computational elements. They do computation, then do some action, reaction, and show intelligent behaviors, so their complexity will be high.
  At design level, it should consider that user should unaware of complexity.
- Agents have to communicate each other and it should check that the action of one agent should not affect the other's action. Side effects should remove carefully, if not possible, put explicit condition to have system stable. This is known as cooperation.
- Agents communicate each other to pass information, so it should be that all agents should use common interface system to communicate, or use a protocol so that they can understand each other clearly.
- Agents should work in the system limitation. It should not be work outside the systems boundaries. It should check all system limitation and try to fulfill its goal that is system dependence.
- Agents should be able to work in group of one or more. The group behavior is critical to perform because all intelligent agents have own view on each decision, so it should be defined properly.

Above it there are many other issues that can be application specific to a particular domain and designers want to add them. These issues can be added as a sub issue or as independent part. Other issues can be mobility, knowledge level, run time behavior, security, privacy, performance etc., as shown in Fig 2. Mobility is with respect to location so we have to consider the mobility issue here. Knowledge level for different agents can be different or same or it can be a part of autonomy. This property defines the system intelligence and help to achieve the goal. Security is very important aspect and very critical issue that helps to make systems invulnerable.

Privacy is another important issue that gives the privacy to agents but in some systems it can create problems, it should be application specific. Performance has the scalability and time response issue that can be measure though simulation of the system or modeling. Performance is very important in every system but critical in Real Time systems design.

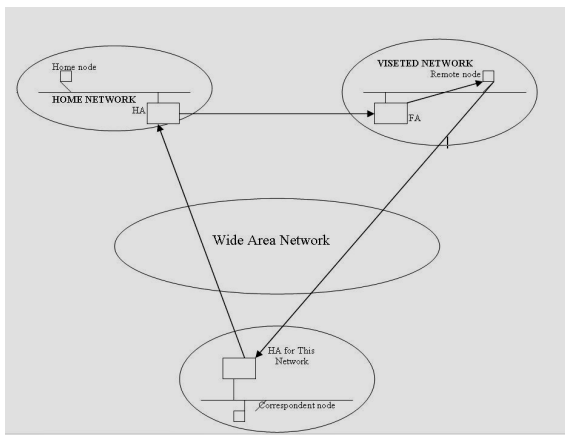

**Figure 3. Mobile Agent Working Scenario**

## CASE STUDY
## ROUTING IN AD-HOC NETWORKS

In this section we are taking the routing scenario on nodes where they are free to relocate. These computing nodes are visiting to other networks and still able to communicate with their local identity [10]. A mobile node can be anything, from a laptop, a mobile phone, a PDA, a smart phone to an iPod also. It have its home configuration and it is pre registered to its HOME ALLOCATION REGISTER where its home agent(HA) is working, that is an intelligent agent and have self learning capability. On the other network the other Agent (which will be foreign agent (FA) for the other network nodes) is working on FOREIGN ALLOCATION REGISTER. Now when the node from home network visits to the other network then FA detects the node that there is a remote node is in my network. It broadcast the address of this node to all and the home agent comes to know that my node is in the network of that particular agent. Now FA sends the information of network address to HA. HA comes to know where its node is and on what address it is operating. Now, when any nodes from outside these two networks want to communicate with that node, so as it, say correspondent node, knows that it should be in its home network, it sends messages to the home address. The HA gets the message at home network, read it and check the status of the node in its register. Now as it is in the remote network and it have the new address, it wraps the message with the new address and send it to FA of that network where node resides. FA reads the message and sees the internal address for the node which is residing in its network, so it sends this message to it. The message has sent to its receiver. If this remote node want to reply for the message to the sender, then it need not to be to go all the way back, as it International Journal of Advanced Engineering Technologyknows the address of the node which sent it that message, it directly sends it the message with its original address as shown in Fig 3. If the correspondent node gets the message, it will get the address of home network because the mobile node sent the message directly, without interference of FA. So it will remain close to correspondent node that where the node is. The correspondent node will think that the node is in its home network only. Here the intelligence of the agents has a great role in all operations. The agents at the design level should be modeled as the interactive, cooperative, autonomous and able to learn. Agents interact to each other and also to other nodes and they do different kind of functions that is very important and difficult to think at design level. The cooperation of agent with each other should be defined priory and simulate them. As it can be that agents are working well alone but in groups they are showing different behavior. Reactivity and Proactivity are the other important issues that are very critical in this example. Reactivity is the phenomenon that it perceives something from the environment and reacts to it. The reaction can be predefined or it can perform on its own. The reactivity should be according of the Proactivity. Proactivity is main goal of the system for which, all agents are working.

The reactivity of an agent can not be known priory, it's a function of artificial intelligence that what it decide at that time. At design time may be its not possible to address all these issue related to reactivity before implementation, But the Proactivity of a system should be clear and well defined.

**CONCLUSION AND FUTURE WORK**

In this paper we discuss the issues that have very important roles in the designing of agent based large concurrent intelligent systems. This paper show why agent-based approach is better than other and what are the issues related to it at design level. The Proposed approach is applicable to all domains and is not dedicated to any particular system or implementation details yet it is very clear to apply. The data represent is based on analysis of agent-based systems and propose the difficulty that generally comes after the design. This paper gives the view of whole system consideration to system designer so that the problems and critical points could be managed and designed carefully, and it helps in designing better Intelligent Systems. We present mobile node example that have many issues involved. Agent-based techniques have many advantages over other design techniques and give better results. This kind of systems would be much helpful for military and other places where intelligent alert is continuously required. Our future work will be based on test and resolve these issue in different domains and refine their specific problems.

**REFERENCES**

1. H. Xu, S. M. Shatz."A Framework forModel-Based Design of Agent-Oriented Software", IEEE Transactions on software engineering, Vol. 29, No.1, USA, Jan 2003.
2. M. Luck, M.d`Inverno,"A Formal Framework for Agency and Autonomy", Procedings of first International conference on Multi-Agent Systems(ICMAS), pp-254-260,1995.
3. M. Fisher, "Representing and executing Agent Based Systems", Intelligent Agents- Proceddings of International workshop on Agent Theories, Arhcitetures and languages, pp 307-323,1995.
4. Amund Tveit:"A survey of Agent-Oriented Software Engineering", First NTNU CSGSC, May, 2001.
5. V. Hilaire, A. Koukam, P. Gruer, and J.-P. Mu¨ller, "Formal Specification and Prototyping of Multi-Agent Systems," Engineering Societies in the Agent World, First International Workshop (ESAW), Aug.2000.
6. D, Kinny, M, Georgeff and A. Rao:"A Methodology and Modelling Technique for Systems of BDI Agents", Technical notes, MAAMAW'96, Springer, Melbourne, Australia, 1996.
7. L. Sterling, T. Juan: "The Software Engineering of Agent-Based Intelligent Adaptive Systems", ICSE'05, St. Louis, Missouri, USA, May15-21, 2005.
8. J. Khallouf, M. Winikoff:" Towards Goal-Oriented Design of Agent Systems", QSIC'05, IEEE, 2005.
9. K. Chan and L. Sterling: "Specifying Roles within Agent-Oriented Software Engineering", 10th IEEE Asia-Pacific Software Engineering Conference (APSE'03), 2003.
10. Kurose James F.," Computer Networking: A Top-Down Approach Featuring the Internet", third Edition, pp 566-570, Pearson Education, 2005.
11. P. Bresciani, et. al.:"A Knowledge Level Software Engineering Methodology for Agent Oriented Programming", AGENTS'01, Quebec, Canada, 28 May-1 June, 2001.
12. K. Chan, L. Sterling, S. Karunasekera:" Agent-Oriented Software Analysis", Australian Software Engineering Conference (ASWEC'04), IEEE, Australia, 2004.
13. T. Juan, A. Pearce, L. Sterling: "ROADMAP: Extending the Gaia Methodology for Complex Open Systems", AAMAS'02, Bologna, Italy, July 2002.